\documentclass[aps,prl,twocolumn,showpacs,superscriptaddress,floatfix]{revtex4}  

\usepackage{graphicx}

\newcommand{\Eqs}[2]{Eqs.~(\ref{#1}) and~(\ref{#2})}

\newcommand{\Fi}[1]{Fig.~\ref{#1}}
\newcommand{\Ta}[1]{Table~\ref{#1}}

\newcommand{\agev}{\mbox{$A$~GeV}}               
\newcommand{\gevc}{\mbox{GeV$/c$}}
\newcommand{\gevcc}{\mbox{GeV$/c^2$}}

\newcommand{\mevcc}{\mbox{MeV$/c^2$}}

\newcommand{\rb}[1]{\mbox{\textrm{\scriptsize #1}}}

\newcommand{\lam}{\ensuremath{\Lambda}}
\newcommand{\lab}{\ensuremath{\bar{\Lambda}}}  

\newcommand{\piplus}{\ensuremath{\pi^+}}
\newcommand{\kmin}{\ensuremath{\textrm{K}^-}}
\newcommand{\kplus}{\ensuremath{\textrm{K}^+}}
\newcommand{\ommin}{\ensuremath{\Omega^-}}
\newcommand{\omminbf}{\ensuremath{\mathbf{\Omega^-}}}   
\newcommand{\omplus}{\ensuremath{\bar{\Omega}^+}}             
\newcommand{\omplusbf}{\ensuremath{\mathbf{\bar{\Omega}^+}}}  
\newcommand{\pt}{\ensuremath{p_{\rb{t}}}}
\newcommand{\mt}{\ensuremath{m_{\rb{t}}}}
\newcommand{\dedx}{\ensuremath{\textrm{d}E/\textrm{d}x}}
\newcommand{\dndy}{\ensuremath{\textrm{d}N/\textrm{d}y}}
\newcommand{\nwound}{\ensuremath{\langle N_{\rb{w}} \rangle}}
\newcommand{\navg}{\ensuremath{\langle N \rangle}}
\newcommand{\tf}{\ensuremath{T_{\rb{f}}}}

\newcommand{\betat}{\ensuremath{\beta_{\rb{t}}}}
\newcommand{\btavg}{\ensuremath{\langle \beta_{\rb{t}} \rangle}}
\newcommand{\betas}{\ensuremath{\beta_{\rb{s}}}}
\newcommand{\der}{\ensuremath{\textrm{d}}}
\newcommand{\omavg}{\ensuremath{\langle \Omega \rangle}}
\newcommand{\piavg}{\ensuremath{\langle \pi \rangle}}
\newcommand{\kpavg}{\ensuremath{\langle \textrm{K}^+ \rangle}}

\newcommand{\tpm}{\ensuremath{\! \pm \!}}
\newcommand{\jpsi}{\ensuremath{\textrm{J}/\psi}}
\newcommand{\psip}{\ensuremath{\psi^{\prime}}}

\begin{document}




\title{\omminbf\ and \omplusbf\ production in central Pb+Pb
collisions at 40 and 158\agev}





\affiliation{NIKHEF, Amsterdam, Netherlands.}  
\affiliation{Department of Physics, University of Athens, Athens, Greece.}
\affiliation{Comenius University, Bratislava, Slovakia.}
\affiliation{KFKI Research Institute for Particle and Nuclear Physics,
             Budapest, Hungary.}
\affiliation{MIT, Cambridge, USA.}
\affiliation{Institute of Nuclear Physics, Cracow, Poland.}
\affiliation{Gesellschaft f\"{u}r Schwerionenforschung (GSI),
             Darmstadt, Germany.} 
\affiliation{Joint Institute for Nuclear Research, Dubna, Russia.}
\affiliation{Fachbereich Physik der Universit\"{a}t, Frankfurt, Germany.}
\affiliation{CERN, Geneva, Switzerland.}
\affiliation{University of Houston, Houston, TX, USA.}
\affiliation{Institute of Physics \'Swi{\,e}tokrzyska Academy, Kielce, Poland.}
\affiliation{Fachbereich Physik der Universit\"{a}t, Marburg, Germany.}
\affiliation{Max-Planck-Institut f\"{u}r Physik, Munich, Germany.}
\affiliation{Institute of Particle and Nuclear Physics, Charles
             University, Prague, Czech Republic.}
\affiliation{Department of Physics, Pusan National University, Pusan,
             Republic of Korea.} 
\affiliation{Nuclear Physics Laboratory, University of Washington,
             Seattle, WA, USA.} 
\affiliation{Atomic Physics Department, Sofia University St.~Kliment
             Ohridski, Sofia, Bulgaria.} 
\affiliation{Institute for Nuclear Studies, Warsaw, Poland.}
\affiliation{Institute for Experimental Physics, University of Warsaw,
             Warsaw, Poland.} 
\affiliation{Rudjer Boskovic Institute, Zagreb, Croatia.}


\author{C.~Alt}
\affiliation{Fachbereich Physik der Universit\"{a}t, Frankfurt, Germany.}
\author{T.~Anticic} 
\affiliation{Rudjer Boskovic Institute, Zagreb, Croatia.}
\author{B.~Baatar}
\affiliation{Joint Institute for Nuclear Research, Dubna, Russia.}
\author{D.~Barna}
\affiliation{KFKI Research Institute for Particle and Nuclear Physics,
             Budapest, Hungary.} 
\author{J.~Bartke}
\affiliation{Institute of Nuclear Physics, Cracow, Poland.}
\author{L.~Betev}
\affiliation{Fachbereich Physik der Universit\"{a}t, Frankfurt, Germany.}
\affiliation{CERN, Geneva, Switzerland.}
\author{H.~Bia{\l}\-kowska} 
\affiliation{Institute for Nuclear Studies, Warsaw, Poland.}
\author{A.~Billmeier}
\affiliation{Fachbereich Physik der Universit\"{a}t, Frankfurt, Germany.}
\author{C.~Blume}
\affiliation{Fachbereich Physik der Universit\"{a}t, Frankfurt, Germany.}
\author{B.~Boimska}
\affiliation{Institute for Nuclear Studies, Warsaw, Poland.}
% ...


\collaboration{The NA49 collaboration}
\noaffiliation



\begin{abstract}
Results are presented on $\Omega$ production in central Pb+Pb 
collisions at 40 and 158\agev\ beam energy. 
For the first time in heavy ion reactions, rapidity distributions 
and total yields were measured for the sum \ommin+\omplus\ at 40\agev\ 
and for \ommin\ and \omplus\ separately at 158\agev.
The yields are strongly underpredicted by the string-hadronic 
UrQMD model but agree better with predictions from hadron gas models.
\end{abstract}



\maketitle

The measurement of multi-strange particles is of particular interest
in heavy ion collisions at ultra-relativistic energies. One important
aspect is the observation that the inverse slope parameter $T$ of the
$\Omega$ \mt spectrum~\cite{WA97MT} is significantly smaller
than expected from the linear mass dependence of $T$ naively
implied by the presence of radial flow.  This led to the hypothesis
that multi-strange hyperons are not affected by the pressure generated
by the hadronic matter in later stages of the reaction~\cite{NUXU}.
Originally, the increase of the production of multi-strange particles
as compared to elementary hadron-hadron collisions was suggested
as a signature of quark-gluon plasma formation~\cite{RAFELS}.
However, existing experimental data on $\Xi$ and \lam\ production at
lower beam energies \cite{NA49LP,E895XI} exhibit a much stronger
enhancement than observed at top SPS energies.  Generally,
it is found that the abundances of strange particles are close to those 
calculated in statistical models assuming the creation of an equilibrated
hadron gas \cite{BECAT1}.  In a hadronic environment, as expected at
lower beam energies, this equilibration is generally difficult to
achieve. At larger energy densities, when the hadronic system might 
be close to the QGP phase boundary, multi-particle fusion processes 
could lead to fast equilibration \cite{PBMJS2}.  However, there
exists no dynamic explanation in a hadronic scenario at lower energy 
densities.  The present measurement of $\Omega$ at 40\agev\ 
provides an important test for these models.  Recent
results on the energy dependence of the ratio 
$\kpavg / \piavg$ \cite{NA49KP,NA49QM} 
indicate a sharp maximum of relative
strangeness production at a beam energy of 30\agev. This observation
can be interpreted as a signal for the onset of deconfinement
\cite{SMES} and might be reflected in the energy dependence of
multi-strange particle production.

The data were taken with the NA49 large acceptance hadron spectrometer
at the CERN SPS. With this detector, tracking is performed by four
large-volume TPCs. A measurement of the specific energy loss \dedx\
provides particle identification at forward rapidities. Time-of-flight
detectors improve the particle identification at mid-rapidity.
Centrality selection is based on a measurement of the energy deposited
in a forward calorimeter by the projectile spectators.  A detailed
description of the apparatus can be found in~\cite{NA49NM}.
  
We present in this paper an analysis of two samples of central Pb+Pb
events taken at beam energies of 40 and 158\agev\ in the years 1999
and 2000, respectively. About $5.8 \times 10^{5}$~events were recorded
at 40\agev\ with a centrality selection of 7.2\% of the total
inelastic cross section corresponding, on average, to $\nwound = 349$
wounded nucleons~\cite{BIALAS}. At 158\agev, $2.8 \times
10^{6}$~events were taken at 23.5\% centrality corresponding to $\nwound
= 262$.

The $\Omega$ were identified in the decay channel \mbox{$\Omega
\rightarrow \lam \textrm{K}$}, \mbox{$\lam \rightarrow \textrm{p} \pi$}
(68\% branching fraction). To reconstruct the \ommin\ (\omplus), the
\lam\ (\lab) candidates were selected in an invariant-mass window of
1.101--1.131~\gevcc\ and combined with all negatively (positively)
charged tracks in the event.
The same procedure as in the $\Xi$ analysis of~\cite{NA49XI} was used
to identify the secondary vertex of the $\Omega$ decay.

To reduce the combinatorial background several cuts were applied.
Identification of the (anti-)protons by \dedx\ in the
TPCs reduced the contribution from fake \lam\ (\lab). The measured
\dedx\ was required to be within 3.5 standard deviations
from the predicted Bethe-Bloch value.  Likewise an enriched kaon
sample was extracted from the charged tracks. A further background
reduction was achieved by requiring a minimal distance of 25~cm in the
beam direction between the target and the $\Omega$ decay vertex
position.  The $\Omega$ candidates were extrapolated back to the
target plane to obtain the transverse coordinates $b_x$ (magnetic
bending plane) and $b_y$ of the impact point with respect to the
primary interaction vertex. To reject non-vertex candidates, cuts of
$|b_x| < 0.5$~cm and $|b_y| < 0.25$~cm were applied. 
Kaons from the primary vertex were excluded by imposing a cut of 
$|b_y| > 1.0$~cm on the kaon tracks. 
In addition, $|b_y| > 0.4$~cm was required for the
$\Lambda$ candidates at 40\agev.  With these cuts an acceptable
separation of signal and background was achieved for $\Omega$
transverse momenta above 0.9~\gevc.

In \Fi{fig:omega} 
\begin{figure}[htb]
\includegraphics[width=0.95\linewidth]{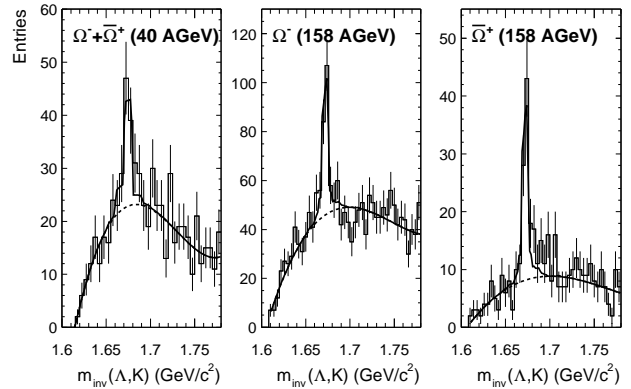}
\caption{\label{fig:omega} 
  The invariant-mass distributions of \ommin\ and \omplus\
  candidates. Left: summed distribution of \lam\kmin\ and \lab\kplus\
  pairs at 40\agev. Middle: \lam\kmin\ pairs at 158\agev. Right:
  \lab\kplus\ pairs at 158\agev.  The full curves represent a fit to
  signal and background described in the text. The dashed curves show
  the background contribution.}
\end{figure}
the invariant-mass distributions of the \ommin\ and \omplus\
candidates are shown for $\pt > 0.9$~GeV/$c$ and $-0.5 < y < 0.5$.
Note that the available statistics at 40\agev\ is not
sufficient to separately analyze the \ommin\ and \omplus\
\cite{MICHI}.  Clear signals are observed at the $\Omega$ mass of 
$m_0 = 1672.5$~\mevcc~\cite{PDG02} with a resolution of 5 and 4~\mevcc\ 
at 40 and 158\agev, respectively.

The spectra were fitted to the sum of a polynomial background and a
signal distribution, determined from the simulation described
below. The raw $\Omega$ yield is obtained by subtracting the fitted
background in a mass window of $\pm 7$~\mevcc\ around the nominal
$\Omega$ mass.

Detailed simulations were made to correct the yields for geometrical
acceptance and losses in the reconstruction. For this purpose, a
sample of $\Omega$ was generated in the full phase space accessible to
the experiment. The Geant~3.21 package \cite{GEANT3} was used to track
the generated $\Omega$ and their decay particles through a detailed
description of the NA49 detector geometry.  NA49 specific software 
was used to simulate the TPC response taking into account all known
detector effects. The simulated signals were added to those of real
events and subjected to the same reconstruction procedure as the
experimental data. The acceptance and efficiency were calculated in
bins of \pt\ and $y$ as the fraction of the generated $\Omega$ which
traverse the detector, survive the reconstruction and pass the
analysis cuts.  This fraction amounts in total to 0.4--0.5\% 
(0.2--0.5\%) for 40 (158)\agev\ data, depending on \pt\ and $y$.
The geometric acceptance is of the order of 20\%, which in turn 
is reduced to 0.6--1.2\% by the cuts that suppress the combinatorial 
background. A further reduction of the reconstruction efficiency at 
158\agev\ by 30--60\% is due to the high track density.

The statistical error is given by the quadratic sum of three 
contributions: the signal, the background and the 
efficiency correction, the latter being smaller at 40 than at 158\agev.
The systematic uncertainties are dominated by the background
subtraction method and by imperfections in the simulation. By varying
the analysis strategy and the cuts applied, a systematic error of 10\%
is estimated in the transverse-mass region $(\mt - m_0) >
0.3$~GeV. At lower \mt\ this error is about 25\%

In \Fi{fig:omegamt}
\begin{figure}[htb]
\includegraphics[width=0.95\linewidth]{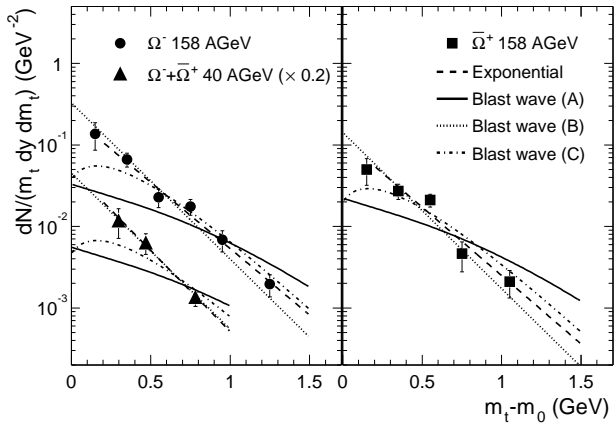}
\caption{\label{fig:omegamt} 
  The transverse-mass spectra of \ommin\ and \omplus\ at mid-rapidity.
  Left: $\ommin + \omplus$ at 40\agev\ (triangles) and \ommin\ at
  158\agev\ (circles). Right: \omplus\ at 158\agev. The errors shown
  are statistical only. The dashed curve shows the exponential fit
  described in the text. The full, dotted, and dash-dotted curves show 
  a model including transverse expansion.}
\end{figure}
the transverse-mass spectra of the $\Omega$ are shown integrated over
the range $\pm 0.5$ (40\agev) and $\pm 1$ (158\agev) around
mid-rapidity.  The transverse-mass spectra were fitted to an
exponential function
\begin{equation}
  \label{eq:expo}
  \frac{\der N}{\mt \der \mt \der y} 
  \propto \exp \left( -\frac{\mt}{T} \right)
\end{equation}
in the range $(\mt - m_0) > 0.2$~GeV. The results are plotted in
\Fi{fig:omegamt} (dashed curves) and the inverse slope parameters $T$
are listed in \Ta{tab:results}.
\begin{table}[tbh]
\caption{\label{tab:results}
  The inverse slope parameter $T$ (MeV), the width $\sigma$ of the
  rapidity distribution, the mid-rapidity yield \dndy\ and the total
  yield \navg\ of $\Omega$ production at 40 and 158\agev. The first
  error is statistical and the second systematic.}
\begin{ruledtabular}
\begin{tabular}{llll}
        &\multicolumn{1}{c}{40\agev}            
        &\multicolumn{1}{c}{158\agev}
        &\multicolumn{1}{c}{158\agev} \\
        &\multicolumn{1}{c}{$\ommin + \omplus$}
        &\multicolumn{1}{c}{\ommin}
        &\multicolumn{1}{c}{\omplus}  \\
\hline
$T$     & $218  \tpm 39   \tpm 39  $ & $267  \tpm 26   \tpm 10  $ 
        & $259  \tpm 35   \tpm 18  $  \\
$\sigma$& $0.6  \tpm 0.1  \tpm 0.1 $ & $1.2  \tpm 0.4  \tpm 0.2 $
        & $1.0  \tpm 0.4  \tpm 0.2 $  \\
$\dndy$ & $0.10 \tpm 0.02 \tpm 0.02$ & $0.14 \tpm 0.03 \tpm 0.01$
        & $0.07 \tpm 0.02 \tpm 0.01$  \\ 
$\navg$ & $0.14 \tpm 0.03 \tpm 0.04$ & $0.43 \tpm 0.09 \tpm 0.03$ 
        & $0.19 \tpm 0.04 \tpm 0.02$ 
\end{tabular}
\end{ruledtabular}
\end{table}
No significant difference can be observed in the shape of the \ommin\
and \omplus\ spectra at 158\agev. The inverse slope parameters are
close to the values obtained by the WA97 and NA57
collaborations~\cite{WA97MT,NA57MT}. The inverse slope parameter at 
40\agev\ is somewhat lower but, within errors, compatible with the 
158\agev\ result.

To investigate whether the $\Omega$ decouples earlier from the
fireball than lighter hadrons, we use a hydrodynamical model which
assumes a transversely expanding emission source \cite{BLASTW}. The
parameters of this model are the freeze-out temperature \tf\ and the
transverse flow velocity \betas\ at the surface.  Assuming a linear
radial velocity profile $\betat (r) = \betas\; r/R$, which is
motivated by hydrodynamical calculations, the \mt spectrum can be
computed from
\begin{equation}
  \label{eq:blast}
  \!\! \frac{\der N}{\mt \der \mt \der y} 
  \propto \! \int_{0}^{R} \!\!\! r \der r\; \mt
  I_{0} \!\!\left(\!\frac{\pt \sinh \rho}{\tf} \!\right)\!
  K_{1} \!\!\left(\!\frac{\mt \cosh \rho}{\tf} \!\right),
\end{equation}
where $R$ is the radius of the source and $\rho = \tanh^{-1}\! \betat$
is the boost angle.  The full curve (A) in \Fi{fig:omegamt} shows the
result of a calculation with $\tf = 90$~MeV and an average flow
velocity $\btavg = 0.5$. These parameters were obtained from a
simultaneous fit of the model to the \mt spectra of \kplus, \kmin, p,
$\bar{\textrm{p}}$, $\phi$, \lam, and \lab, all measured by NA49 at
158\agev~\cite{NA49KP,NA49PR,NA49PH,NA49LP}. The dotted curve (B) is
calculated with $\tf = 170$~MeV and $\btavg = 0.2$, obtained from a
fit to \jpsi\ and \psip\ spectra~\cite{MAREK1}.  The disagreement 
of curve (A) and the agreement of curve (B) with the data
suggest that in this version of the  model, the freeze-out conditions 
of the $\Omega$ are similar to those of the \jpsi\ or \psip\ 
but are different from those of the lighter hadrons. 
The use of a constant expansion velocity, on the other hand, results
in a fair agreement of the measured hadron spectra, including the $\Xi$ 
and the $\Omega$, with $\tf = 127$~MeV and $\btavg = 0.5$~\cite{MARCO}, 
as shown by curve (C) for the Omega.
It predicts, however, a significant decrease of the spectrum for
heavy hadrons for $\mt - m_{\rb{0}} \rightarrow 0$.  This dip is
already quite pronounced for the Omega but not suggested by the
measurements \footnote{Note that the transverse-mass spectra at 
158\agev\ presented here deviate in the first data point from the 
result of a preliminary analysis shown in \cite{MARCO}.}.
Thus, a radius independent transverse expansion 
velocity may be too crude an approximation of the velocity profile. 

The parameterizations of \Eqs{eq:expo}{eq:blast} were used to
extrapolate the $\Omega$ yields into the unmeasured regions of \mt.
Assuming that the shape of the \mt distribution does not depend on
rapidity, extrapolation factors of 2.3~(2.2) at 40~(158)\agev\ were
obtained from fits to the summed \ommin\ and \omplus\ data. A
systematic uncertainty of 6\% is due to the choice of
parameterization.

The extrapolated $\Omega$ yields at 40 and 158\agev\ are shown in
\Fi{fig:omegarap} as a function of rapidity.
\begin{figure}[htb]
\includegraphics[width=0.95\linewidth]{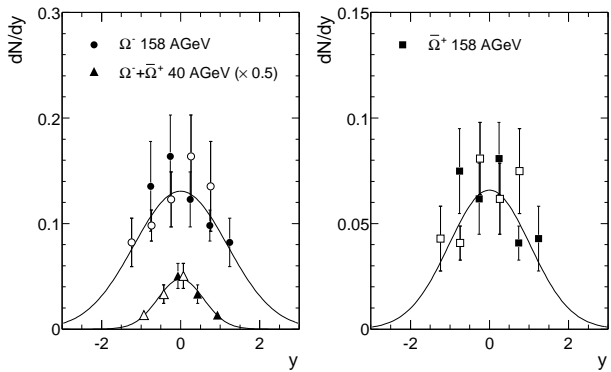}
\caption{\label{fig:omegarap}
  The rapidity dependence of $\Omega$ production in central Pb+Pb
  collisions. Left: $\ommin + \omplus$ at 40\agev\ (triangles) and
  \ommin\ at 158\agev\ (circles). Right: \omplus\ at 158\agev. The
  errors shown are statistical only. The open symbols show the
  measured points (full symbols) reflected around mid-rapidity. The
  curves correspond to Gaussian fits to the data.}
\end{figure}
All spectra can be described by a Gaussian with zero mean and a width
$\sigma$ obtained from a fit to the data, see
\Ta{tab:results}. 
The widths of the \ommin\ and \omplus\ spectra at 158\agev\ are
compatible but are both significantly larger than the width measured
at 40\agev.
Also given in \Ta{tab:results} are the mid-rapidity yields \dndy\
($-0.5 < y < 0.5$) and the total yields \navg\ obtained by
extrapolating the rapidity spectra into the unmeasured region using
the Gaussian fits.
The mid-rapidity yields are slightly below the values given by the NA57 
collaboration~\cite{NA57DN}, however, they agree within 
statistical errors, if the difference in the centrality selection at 
158\agev\ is taken into account.

\begin{figure}[htb]
\includegraphics[width=0.85\linewidth]{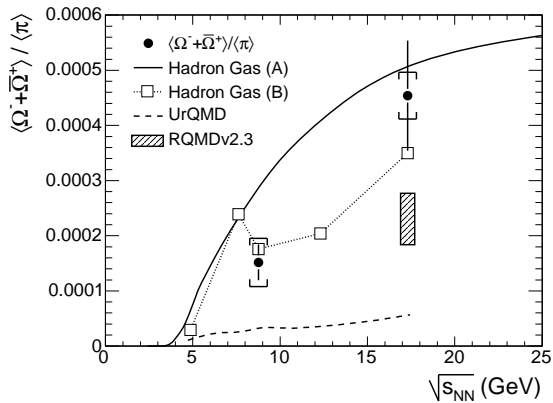}
\caption{\label{fig:omegaedep} 
  The ratio $\omavg / \piavg$ (see text) versus the center-of-mass
  energy. Statistical errors are shown as lines, while the brackets
  denote the systematic errors.  The dashed curve shows the prediction 
  from the hadronic string model UrQMD~\cite{URQMD} and the gray box 
  that of RQMD~\cite{RQMD}. A hadron gas model without strangeness
  suppression~\cite{REDLIC} is shown by the full curve.  The open
  squares represent the fits from~\cite{BECAT2} including strangeness
  under-saturation.}
\end{figure}

In the following we denote by \omavg\ the sum of the total \ommin\ and
\omplus\ yields and by \piavg\ the total charged pion yields
from~\cite{NA49KP}, multiplied by a factor 1.5.
The pion yields at 158\agev\ were scaled by the ratio of the numbers of 
wounded nucleons to account for the difference
in the centrality selection of the pion and the $\Omega$ measurement
(note the centrality selection at 40\agev\ of 7\% and at 158\agev\ 
of 23.5\% most central events.).
Figure~\ref{fig:omegaedep}
shows the ratio $\omavg / \piavg$ as function of the center-of-mass
energy.  The ratio tends to increase with energy.
It is clearly underpredicted by the UrQMD string-hadronic
model~\cite{URQMD} as shown by the dashed curve in
\Fi{fig:omegaedep}. A better description of the 158\agev\ data is
provided by RQMD version 2.3 including the color rope
mechanism~\cite{RQMD}.

On the other hand, the data are close to the predictions of
statistical hadron gas models which use a grand canonical ensemble.
In these models, the chemical freeze-out temperature and the baryonic
chemical potential are fitted
to the yields of other measured hadrons.  The hadron gas model
of~\cite{BECAT2} (labeled B in \Fi{fig:omegaedep}) introduces in
addition a strangeness undersaturation parameter $\gamma_{\rb{s}}$
in the fits, which have been performed at each energy 
separately.  The fit results at 30\agev\ reflect the sharp maximum of 
the \kplus/\piplus-ratio observed around this energy.  The present 
measurement at 40\agev\ seems to favor this model, compared to that 
of~\cite{REDLIC} (labeled A in \Fi{fig:omegaedep}) which does not allow 
for strangeness undersaturation ($\gamma_{\rb{s}} = 1$ for all energies). 
The data point at 158\agev, however, does not discriminate between the two
models.  Nevertheless, the observation that $\Omega$ production is 
compatible with phase-space undersaturation at higher SPS energies 
($> 30\agev$), would be in line with a similar behavior of the kaon 
excitation function in the same energy regime~\cite{NA49QM}.


In summary, NA49 has performed a measurement of $\Omega$ production in
central Pb+Pb reactions over a wide region of phase space. At a beam
energy of 158\agev\ the available statistics allowed to separately
analyze \ommin\ and \omplus.  The shapes of the transverse-mass
spectra at this energy reveal no difference between \ommin\ and
\omplus\ and are in agreement with previous results by WA97 and NA57.  
In a hydrodynamically inspired  model with radially increasing
velocity profile, the data favor a low transverse expansion velocity 
and high freeze-out temperature.  The rapidity spectra of the $\Omega$, 
which have not been measured before in heavy ion reactions, 
are compatible with a Gaussian shape. The widths for \ommin\ and 
\omplus\ appear to be similar. The yields are strongly under-predicted 
by the string-hadronic UrQMD model.  The data agree better with 
predictions from hadron gas models.


\begin{acknowledgments}
Acknowledgments: This work was supported by the US Department of
Energy Grant DE-FG03-97ER41020/A000, the Bundesministerium f\"ur Bildung
und Forschung and the Virtual Institute VI-146 of the Helmholtz 
Gemeinschaft, Germany, the Polish State Committee for Scientific
Research (2 P03B 130 23, SPB/CERN/P-03/Dz 446/2002-2004, 2 P03B
04123), the Hungarian Scientific Research Fund, OTKA (T032648,
T032293, T043514, F034707), the Polish-German Foundation, and the
Korea Research Foundation Grant (KRF-2003-070-C00015).
\end{acknowledgments}




\end{document}